\documentclass[aps,prl,twocolumn,groupedaddress,showpacs]{revtex4}
\usepackage{graphicx}
\usepackage{amssymb}

\begin{document}


\title{Projectile-shape dependence of impact craters in loose
granular media}


\author{K.A. Newhall and D.J. Durian}
\affiliation{UCLA Department of Physics \& Astronomy, Los Angeles, CA
90095-1547}


\date{\today}

\begin{abstract}
    We report on the penetration of cylindrical projectiles dropped
    from rest into a dry, noncohesive granular medium.  The cylinder
    length, diameter, density, and tip shape are all explicitly
    varied.  For deep penetrations, as compared to the cylinder
    diameter, the data collapse onto a single scaling law that varies
    as the 1/3 power of the total drop distance, the 1/2 power of
    cylinder length, and the 1/6 power of cylinder diameter.  For
    shallow penetrations, the projectile shape plays a crucial role
    with sharper objects penetrating deeper.
\end{abstract}



\maketitle



Anyone who has walked across a beach realizes that even though the
sand moves under the pressure of a step, at some point their foot is
supported.  Someone who is running leaves a deeper and larger
footprint, or crater, than someone who is walking.  This phenomenon
applies to anything dropped into sand; no matter how energetic the
impact, the sand will eventually stop the projectile.  For an object
of mass $m$ that is dropped from rest and that moves downwards a total
distance $H$ (including motion throughout the impact), the
gravitational potential energy $mgH$ is transfered from the projectile
to the sand.  Some of this energy is dissipated and some of it is
converted back to gravitational potential energy associated with the
size and shape of the crater.  Fundamentally, the energy is expended
by an average stopping force $\langle F\rangle$ applied by the sand as
the projectile moves a total distance $d$ from initial impact to rest:
\begin{equation}
	\langle F\rangle d = mgH.
\label{mgH}
\end{equation}
Thus, a straightforward measurement of $d$ vs $H$ provides information
about granular mechanics, a topic of widespread
interest~\cite{jnb,duran}.  A full understanding of the stopping
forces and the transfer of energy may enable a fundamental explanation
for such impact-related phenomena as the complex sequence of crater
morphologies~\cite{roddy,melosh,holsapple,deBruyn} and the
spectactular granular jets formed by a collapsing
hole~\cite{siggi,detlef}.

\begin{figure}
\includegraphics[width=3.00in]{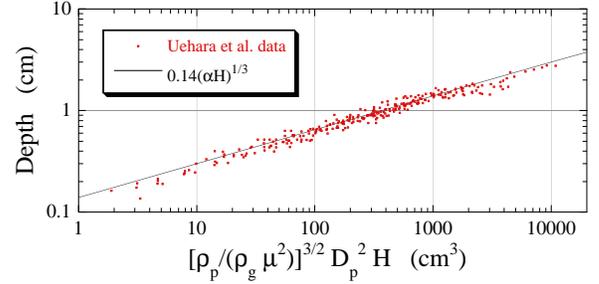}
\caption{Crater depth vs scaled total drop distance.  All data are
from Ref.~\protect{\cite{jun}} for spherical projectiles.  The solid 
line is the power-law scaling given by Eq.~(\protect{\ref{junlaw}}).
\label{jundata}}
\end{figure}

The penetration of spheres into loose granular media has been reported
previously.  In Ref.~\cite{jun}, the total drop distance $H$ was
varied by up to a factor of 1000; the projectile density $\rho_{p}$
was varied by a factor of 60; the projectile diameter $D_{p}$ was
varied by a factor of 4; the grain density $\rho_{g}$ was varied by a
factor of 2; and the tangent of the draining repose angle $\mu$ was
varied by a factor of 2.  The resulting penetration depths are
reproduced in Fig.~\ref{jundata}, along with the empirical scaling law
\begin{equation}
    d = 0.14\left[\left({\rho_{p}\over\rho_{g}\mu^{2}}\right)^{3/2} 
    {D_{p}}^{2}H\right]^{1/3}.
\label{junlaw}
\end{equation}
Evidently, the crater depth does not vary simply with either the
kinetic energy of the sphere, $\sim\rho_{p}{D_{p}}^{3}H$, or its
momentum, $\sim\rho_{p}{D_{p}}^{3}H^{1/2}$.  In Ref.~\cite{deBruyn} a
different definition of crater depth, suitable when the spheres
completely submerge, gave different scaling.

The behavior in Fig.~\ref{jundata} cannot be explained by current
theories.  Granular hydrodynamics and a rate-independent ``plowing''
force both predict different scaling than Eq.~\ref{junlaw} (see
discussion in Ref.~\cite{jun}).  But how general is the observation? 
In particular, how does the crater depth depend upon the shape of the
projectile?  One would expect a pointed projectile to penetrate deeper
than a blunt projectile.  Perhaps this may be accounted for by the
value of the numerical coefficient in Eq.~(\ref{junlaw}).  Or perhaps
the functional form of depth vs drop distance may be entirely
different, with $d\sim H^{1/3}$ arising as an accident of the
projectile's spherical shape.

To explore the role of projectile shape on impact cratering, we
perform a series of experiments with the same basic protocol as in
Ref.~\cite{jun}.  Namely, we fill a 1000~mL beaker (diameter $4''$)
with dry, noncohesive glass beads (diameter 0.2~mm, draining respose
angle $24^{\circ}$).  The depth of filling is $2''$ for most trials,
but is increased to $4''$ for deep impacts.  The density of the
granular medium is $\rho_{g}=1.51$~g/cc, which corresponds to
random-close packing with a volume fraction of about 63\%.  As in
Ref.~\cite{jun}, this state is achieved by swirling the sample
horizontally, at first rapidly and then ever more slowly, until the
surface is level and at rest.  Instead of dropping spheres, we now
drop a wide variety of cylinders and measure their depth of
penetration.  Cylinder materials include wood (poplar,
$\rho_{p}=0.5$~g/cc), Aluminum (Al, $\rho_{p}=2.7$~g/cc), and
high-density Tunsten Carbide (WC, $\rho_{p}=17.1$~g/cc).  We use four
different cylinder diameters, $D_{p}\in\{0.25'', 0.50'', 1.0'',
2.0''\}$, and four different cylinder lengths, $L\in\{0.5'', 1'', 2'',
4''\}$.  To vary the projectile shape, progressively sharper conical
tips are machined at one end of these cylinders; we use four different
cone angles, $\{180^{\circ}~{\rm (i.e.\ flat)}, 135^{\circ},
90^{\circ}, 45^{\circ}\}$.  To achieve a vertical orientation at
impact, the cylinders are suspended by a thin loop of thread, tied
through an accurately machined hole.  For each cylinder, the drop
distances are varied as widely as possible within certain limitations. 
The minimum is set either by the penetration depth for a cylinder
placed infinitessimally above the surface of the medium, or by the
tendency of (long or light) cylinders to fall over after a shallow
penetration.  The maximum drop distance is set either by the depth of
the sand or by the tendency of cylinders to begin to tumble during
free-fall.  No trial is accepted unless the cylinder remains vertical
after impact.

\begin{figure}
\includegraphics[width=3.00in]{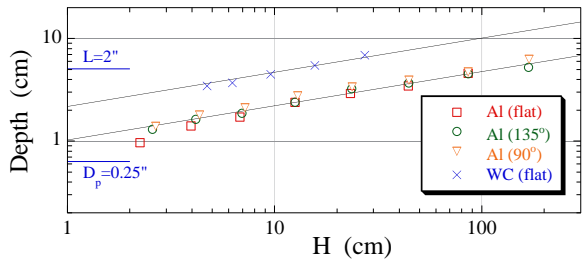}
\caption{Crater depth vs drop distance for four different cylindrical
projectiles, all with diameter $D_{p}=0.25''$ and length $L=2''$.  The
solid lines represent fits to a 1/3 power-law, $d\sim H^{1/3}$.
\label{rawdata}}
\end{figure}

Four sets of penetration depth vs drop distance data are displayed in
Fig.~\ref{rawdata}, all for cylindrical projectiles of length $2''$
and diameter $0.25''$.  This includes WC and Al rods with flat ends,
as well as Al rods with $135^{\circ}$ and $90^{\circ}$ degree tips. 
Evidently WC penetrates deeper than Al, in accord with their
densities.  The three Al rods all penetrate to approximately the same
depth, in spite of their different tip shapes.  Furthermore, all data
are adequately described by a 1/3 power-law, $d\sim H^{1/3}$, just as
for spherical projectiles.  Therefore, the projectile shape seems to
play no crucial role and the prior scaling law is not special to
spheres.

\begin{figure}
\includegraphics[width=3.00in]{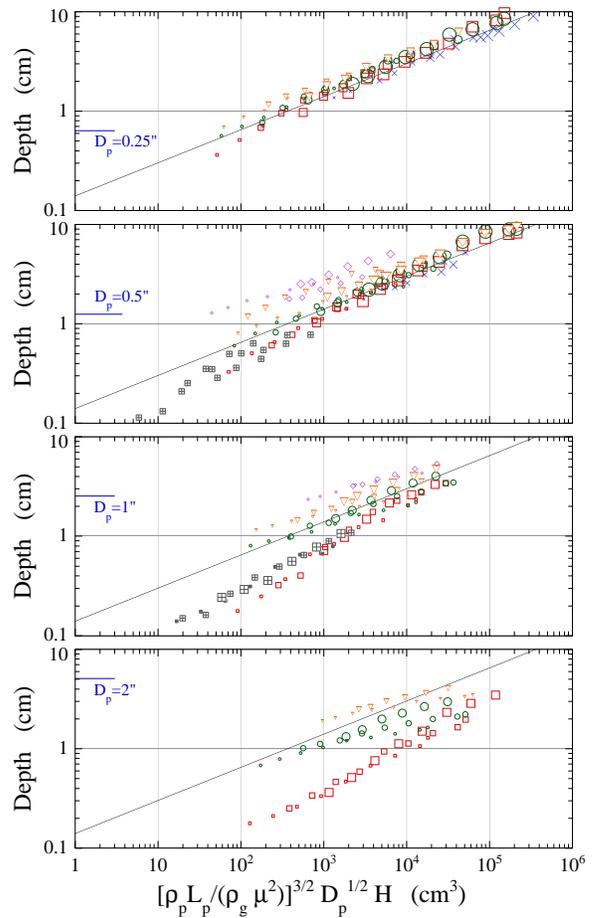}
\caption{Crater depth vs scaled total drop distance.  Data is for
cylinderical objects, divided into graphs by cylinder diameter, as
labelled.  Values of $L$ varried from $0.5''$, indicated by the
smallest points, to $1.0''$, to $2.0''$, to $4''$ by the largest
points.  Tungsten Carbide is denoted by $\times$, flat-ended Aluminum
by $\square$, $135^\circ$-tip Aluminum by $\bigcirc$, $90^\circ$-tip
Aluminum by $\triangledown$, $45^\circ$-tip Aluminum by $\Diamond$ and
flat ended wood by $\boxplus$.  The line in each subplot is
Eq.~(\protect\ref{hypothesis}).
\label{ktdata}}
\end{figure}

The results in Fig.~\ref{rawdata} severely constrain the
generalization of Eq.~(\ref{junlaw}) to aspherical projectiles.  A
natural hypothesis is that the crucial parameter is the mass per unit
cross sectional area, $(2/3)\rho_{p}D_{p}$ for spheres and $\rho_{p}L$
for cylinders.  For cylinders, equal variation of density and length
should lead to equal variation of penetration depth.  Thus, we must
have
\begin{equation}
    d=0.14\left[\left({\rho_{p}L\over\rho_{g}\mu^{2}}\right)^{3/2}
    {D_{p}}^{1/2}H\right]^{1/3},
\label{hypothesis}
\end{equation}
for consistency with Eq.~(\ref{junlaw}).

To test this expectation, we display depth data vs scaled drop height,
$[(\rho_{p}L)/(\rho_{g}\mu^{2})]^{3/2}{D_{p}}^{1/2}H$, in
Figs.~\ref{ktdata}(a-d).  This allows us to plot
Eq.~(\ref{hypothesis}) as a single curve, onto which we look for all
data to collapse.  Each subplot is for a different rod diameter;
symbol types distinguish rods according to their length, density, and
tip shape.  Evidently, the degree of collapse is better for deeper
penetrations and is worse for shallower penetrations.  In
Fig.~\ref{ktdata}a, for the $0.25''$ diameters, the quality of collapse
is comparable to that found earlier for spheres.  Comparison of the
four subplots shows that the onset of collapse appears correlated with
rod diameter.  In particular, collapse by Eq.~(\ref{hypothesis}) seems
to be attained when the penetration is larger than the rod diameter. 
For shallower penetration, Eq.~(\ref{hypothesis}) fails and the actual
depth depends on tip shape.  In this regime, sharper tips cause deeper
penetration.  The tendency of the data to deviate from $d\sim H^{1/3}$
for small $H$, concomitant with the onset of shape dependence, can
also be seen in close re-inspection of Fig.~\ref{rawdata}.

In conclusion, the crater depth vs drop height scaling observed
previously for spheres is robust and can be simply generalized to
projectiles of different shape.  The $d\sim H^{1/3}$ power law is not
an accident of projectile shape.  The only accident is that it holds
for spheres even when the penetration is not greater than the sphere
diameter.  The independence of deep penetrations on projectile shape,
as observed here, should heighten the importance of finding a full
theoretical understanding.  It also offers a clue: The dissipation of
the projectile's energy in the granular medium must occur throughout a
volume that is much larger than the crater.  Presumably, this involves
a cooperative effect of many grains, such as long force chains
successively created then broken by the projectile impact.

To bring this manuscript full circle, we return to a walk on the
beach.  In particular, we generalize the depth scaling law to
projectiles with non-circular cross section and we predict the depth
of footprints in dry noncohesive sand ($\rho_{g}=1.59$~g/cc,
$\mu=\tan(38^\circ)$ \cite{jun}).  Making three substitutions,
$\rho_{p}L$ with the mass per cross sectional area $m/A$, $D_{p}$ with
$2\sqrt{A/\pi}$, and $H$ with $v^{2}/(2g)$, we arrive at
\begin{equation}
    d=0.11 \left({m/A^{3/2}\over\rho_{g}\mu^{2}}\right)^{1/2}
    \left({Av^{2}\over g}\right)^{1/3}.
\label{noncircular}
\end{equation}
The area factors have been distributed to make it more apparent that
this expresssion is dimensionally correct; at constant mass, the depth
decreases with area as $d\sim A^{-5/12}$.  For impact by a foot, we
take $A=8\times30$~cm$^{2}$, $m=70,000$~g, and $v=100$~cm/s
(corresponding to a drop height of 5~cm).  For these numbers
Eq.~(\ref{noncircular}) predicts a footprint depth of 6.7~cm, in
reasonable accord with experience.

{\it Acknowledgements} We thank Fran{\c c}oise Queval for coordinating
the ``Research Experience for Undergraduates'' program at UCLA. This
material is based upon work supported by the National Science
Foundation under grants DMR-0305106 and PHY-0243625.

\bibliography{CraterRefs}

\end{document}